# LA AVENTURA DE LA FÍSICA

Andrey Chubykalo, Valeri Dvoeglazov, Augusto Espinoza y Alejandro Gutiérrez
Facultad de Física, Universidad Autónoma de Zacatecas
Apartado Postal C-580 Zacatecas 98062, Zac., México
e-mail: achub@prodigy.net.mx, agarrido@cantera.reduaz.mx,
agtrrezr@cantera.reduaz.mx, valeri@cantera.reduaz.mx

**Resumen**

Con la intención de realizar una pequeña aportación en la celebración del año 2005, proclamado como Año Internacional de la Física, el cuerpo académico de Partículas, Campos y Astrofísica de la Universidad Autónoma de Zacatecas presenta un pequeño esbozo de los trabajos realizados por Einstein en 1905, así como un conjunto de ensayos sobre electrodinámica clásica, el grupo de Lorentz, neutrinos y relatividad general, todos éstos relacionados con los temas de investigación que realizamos.

## *INTRODUCCIÓN*

El año 2005 ha sido proclamado por las Organización de las Naciones Unidas el Año Internacional de la Física en conmemoración del centenario de la publicación por parte de Albert Einstein de 5 artículos que influyeron considerablemente en nuestra concepción del mundo, y en reconocimiento de la importancia de la Física y de la creación de infraestructura científica como base para el desarrollo tecnológico de la sociedad.

En este artículo que pretende ser un modesto aporte a la gran celebración del Año Internacional de la Física se presentan algunos tópicos relacionados con las investigaciones realizadas en los últimos años por el Cuerpo Académico de Partículas, Campos y Astrofísica de la Unidad Académica de Física de la Universidad Autónoma de Zacatecas. La relación de estas contribuciones con las ideas desarrolladas por Albert Einstein es obvia ya que toda la física contemporánea está de alguna manera vinculada con las aportaciones fundamentales de este gran científico.

Comenzaremos con un breve bosquejo de las publicaciones que son motivo de este homenaje para después ofrecer una serie de pequeños ensayos relativos a los temas que son motivo de nuestro trabajo de investigación.

El 17 de Marzo de 1905 Einstein envía a Annalen der Physik su artículo sobre el efecto fotoeléctrico *"Un punto de vista heurístico sobre la producción y transformación de la luz"*. En este artículo Einstein demostró que es insuficiente describir la luz solamente como un fenómeno ondulatorio, sino que es necesario describirla como si ella consistiera de cuantos de energía independientes. Con esta idea él logra explicar el misterio de cómo la radiación electromagnética logra extraer electrones de los metales por medio del efecto fotoeléctrico. Fue el comienzo de una relación conflictiva con la mecánica cuántica. Por este trabajo Einstein recibió en 1921 el Premio Nóbel de Física

Posteriormente, el 30 de Abril de 1905 Einstein termina su segundo trabajo, *"Una nueva determinación de las dimensiones moleculares"*, en el cual él muestra como calcular el número de Avogadro y las dimensiones de las moléculas estudiando su movimiento en una solución. Este artículo es aceptado como tesis doctoral por la Universidad de Zurich en Julio y publicada, en una variante un poco diferente, por Annalen der Physik en Enero de 1906. A pesar de que la fama de este trabajo ha sido oscurecida por la fama de sus artículos sobre la relatividad espacial y el efecto fotoeléctrico, la tesis de Einstein sobre las dimensiones moleculares se ha convertido en uno de sus trabajos mas citados sobre todo por su relación con la mecánica estadística que forma la base de algunos de sus avances excepcionales, incluyendo la idea de que la luz está cuantizada.

Su artículo *"Acerca del movimiento de pequeñas partículas suspendidas en líquidos estacionarios requiere de la teoría cinético-molecular del calor"* fue enviado a Annales der Physik el 11 de Mayo de 1905. En este artículo Einstein combina la teoría cinética y la hidrodinámica clásica para obtener una ecuación que muestra que el desplazamiento browniano de las partículas varía como la raíz cuadrada del tiempo. Con esto proporcionó el marco teórico para la comprobación experimental de la existencia real de los átomos, cosa importante si recordamos que en aquella época la mayoría de los físicos no creía en los átomos. La comprobación experimental fue realizada por Jean Perrin tres años mas tarde.

En el artículo *"Sobre la electrodinámica de cuerpos en movimiento"*, enviado a la revista Annales der Physik el 30 de Junio de 1905, Einstein introducía la teoría especial de la relatividad que cambio completamente nuestras nociones acerca del espacio y del tiempo. Antes de la publicación de este trabajo se tenía la percepción de que las leyes físicas debían transformarse, con ayuda de las transformaciones de Galileo, de un sistema de referencia a otro de tal modo que estas leyes no dependieran del sistema de referencia. Sin embargo, la electrodinámica desarrollada por Maxwell a finales del siglo XIX planteaba un problema fundamental a este principio de relatividad porque ella sugería que las ondas electromagnéticas siempre viajan a la misma velocidad. Por lo tanto, alguna de estas dos teorías, la electrodinámica de Maxwell o la mecánica de Newton, era incorrecta. Fiel a su estilo de pensamiento Einstein desechó el concepto del éter y planteó la audaz hipótesis de que independiente del movimiento de la fuente o del observador la luz siempre se propaga con la misma velocidad. Esta hipótesis junto con el requerimiento de que las leyes físicas deben ser idénticas en todos los sistemas inerciales de referencia, permitió a Einstein construir una nueva teoría del movimiento que revelaba que la mecánica de Newton era sólo una aproximación válida para velocidades pequeñas. Las consecuencias más extrañas de esta teoría eran que la longitud de un objeto se hace más corta si éste viaja a velocidad constante y que los relojes en movimiento caminan más lentamente que los que están en reposo.

El 27 de Septiembre Einstein envía a *Annales der Physik* el último artículo de este año *"¿Depende la inercia de un cuerpo de la energía contenida en él?"*. Este trabajo fue una consecuencia más, probablemente la más famosa, de su teoría de la relatividad, y se sintetiza en la conocida fórmula $E=mc^2$, la cual permite la conclusión espectacular de que la masa en movimiento y la energía son lo mismo.

Pero 1905 fue sólo el comienzo. La teoría general de la relatividad, el logro mas destacado de Einstein, fue desarrollada 10 años mas tarde, con sus predicciones sobre el desplazamiento del perihelio de Mercurio y la desviación de la luz por cuerpos masivos, la cual, a su vez, fue confirmada algunos años después durante el eclipse solar de 1919. Si la teoría especial de la relatividad nos ayuda a comprender el micromundo de las partículas elementales y sus interacciones, la teoría general de la relatividad revolucionó nuestra concepción del universo al predecir fenómenos astrofísicos tan fantásticos como el Big Bang, las estrellas neutrónicas, los agujeros negros y las ondas gravitacionales.

## *I. LA ELECTRODINÁMICA CLÁSICA SIGUE SIENDO UNA CIENCIA VIVA.*

A principios del siglo XIX se dio una gran importancia a la electricidad y al magnetismo en la tentativa de justificar una perspectiva particular del mundo. Después del descubrimiento fundamental de Faraday de la ley de la inducción, el desafío se presentó en la unificación de la electrodinámica en un cuerpo coherente fuera de la electrostática de Neumann-Coulomb y de la magnetostática de Ampere. Desde el punto de vista histórico era un momento único y excitante en el desarrollo de la física [1]. Dos conceptos rivales, el concepto de la acción instantánea a distancia y el campo de Faraday, estaban a la espera de un progreso decisivo en la teoría y en el experimento para ser confirmados o rechazados. El estado del electromagnetismo se caracterizaba por la búsqueda de un concepto correcto e inequívoco que excluyera todas las alternativas. En 1848 Wilhelm Weber fue el primero que intentó la unificación de la teoría electromagnética [2]. En estos tiempos la teoría de Weber del tipo de acción a distancia era considerada como un gran avance porque reproducía los resultados de Neumann-Coulomb y de Ampere usando un tratamiento analítico de las corrientes inducidas. Algunos años más tarde en 1855-56, James Clerk Maxwell comunicó a la Sociedad Filosófica de Cambridge su primera memoria: una tentativa de desarrollar un concepto mecánico comprensivo del campo electromagnético. Esto se convirtió luego en una poderosa teoría de campo [3,.4]. En los años inmediatamente después del tratado de Maxwell, los experimentos proporcionaron cierta cantidad de evidencia en favor de su teoría. Sin embargo esto no era suficiente para elegir entre las dos alternativas. Las dos teorías, la de Weber y la de Maxwell, podían describir satisfactoriamente la mayoría de los fenómenos electromagnéticos conocidos a pesar de las muchas limitaciones internas.

En 1870, como consecuencia de la larga e infructuosa oposición entre estos dos conceptos rivales en el electromagnetismo, y con el fin de reconciliarlos, Helmholtz [5] propuso una teoría intermedia. Pronto se convirtió en la teoría aceptada en Alemania y Europa continental. En particular, Hertz y Lorentz se familiarizaron con la teoría de Maxwell a través de las ecuaciones de Helmholtz. La teoría intermedia de Helmholtz también era incompleta e internamente inconsistente. Se aplicó exclusivamente a un medio dieléctrico en reposo y no consideraba las corrientes de desplazamiento en el vacío (éter). La teoría de Helmholtz permitió la coexistencia simultánea de modos eléctricos longitudinales instantáneos y de ondas eléctricas y magnéticas transversales que se propagan con una cierta velocidad finita. El campo magnético, sin embargo, se podía transmitir solamente con modos transversales. Había un conflicto conceptual con la teoría del Maxwell. Por ejemplo, en un medio dieléctrico el valor de la velocidad transversal de la onda difería considerablemente de la predicha por las ecuaciones de campo de Maxwell.

En estos momentos había una necesidad urgente de datos experimentales nuevos y confiables. La Academia de Berlín propuso como tema a concurso "Establecer experimentalmente una relación entre la acción electromagnética y la polarización de dieléctricos". Esta investigación condujo a Hertz a un descubrimiento de gran importancia, el cual ahora es considerado por muchos como un momento crucial en la historia del electromagnetismo. Hertz tuvo éxito al observar directamente la propagación de ondas electromagnéticas en el espacio libre con la velocidad predicha por Maxwell, y de este modo comprobó rotundamente su teoría. Las otras dos alternativas (la teoría de Weber y la de Helmholtz) fueron rechazadas a pesar de que la teoría de Helmholtz fue corregida para que la velocidad de propagación para los modos transversales no contradijera el experimento de Hertz. En la literatura dedicada a la historia y a la metodología de la física ha habido una gran discusión de si un experimento como el de Hertz se puede considerar como un argumento decisivo en la elección entre diferentes alternativas [6]. A partir de este momento todas las investigaciones importantes sobre electromagnetismo se basaron en las ecuaciones de Maxwell. Sin embargo, esta teoría todavía adolecía de algunos de los pequeños problemas inherentes a sus precursores. Para ser más específicos, las ecuaciones de Maxwell para procesos estacionarios podían ser compatibles solamente con las corrientes continuas cerradas derivadas de la magnetostática de Ampere. La teoría necesitaba una adecuada generalización al caso de un sistema formado por una sola carga.

El primer análisis de este problema desde el punto de vista de la teoría de Maxwell fue emprendida por J. J. Thompson en 1881 [7] y más adelante por O. Heaviside [8]. Ellos encontraron que un cambio en la localización de la carga (ellos consideraban un cuerpo electrostáticamente cargado en movimiento) debe producir una alteración continua del campo eléctrico en cualquier punto del medio circundante (un punto de diferencia con la teoría de Maxwell de campos estacionarios). En el lenguaje adoptado por Maxwell, debe haber corrientes de desplazamiento en el vacío, atribuidas a los efectos magnéticos de la carga móvil. Desarrollando este enfoque, FitzGerald, en una corta pero valiosa nota [9], precisó que el método de Thompson se debe identificar con la hipótesis básica de la teoría de Maxwell sobre la corriente total. Esta conclusión estaba basada en el supuesto que la carga en movimiento, en si misma, debía ser considerada como un elemento de la corriente. Entonces la corriente total, compuesta por la corriente de desplazamiento y por la de la carga móvil, era circuital, de acuerdo con las ideas fundamentales de Maxwell.

Como resultado de sus investigaciones Thompson y Heaviside encontraron el campo eléctrico y magnético de una carga en movimiento, lo que ahora puede verificarse directamente por medio de las transformaciones de Lorentz. Ellos también hallaron por primera vez una fórmula explicita para la fuerza mecánica que actúa sobre una carga eléctrica que se mueve en un campo magnético, conocida ahora como fuerza de Lorentz. Ésta fue la primera demostración exitosa de que la teoría de Maxwell podía ser extendida a un sistema de una carga.

A pesar de los avances logrados por Maxwell y sus seguidores más tempranos era necesaria una teoría más general para los cuerpos en movimiento. En 1890, Hertz hizo el primer intento de construir esta teoría. Como en todos los métodos del siglo XIX, Hertz basó sus consideraciones en un modelo adecuado del éter. Como Hertz mismo lo expresó, "el éter contenido dentro de los cuerpos ponderables se mueve con ellos". Al igual que Maxwell, él asumió que el estado del sistema compuesto (materia y éter) se podría especificar de la misma manera cuando la materia estaba en movimiento o en reposo. Así, fue Hertz quien por primera vez estableció el principio de la relatividad en la electrodinámica. No habría podido ser tan general como el principio de la relatividad de Einstein-Poincarè puesto que estaba atado al modelo del éter. Sin embargo, como nosotros consideramos [16], las ecuaciones de Hertz, reconsideradas cuidadosamente, se pueden tratar como una forma relativista invariante y generalizada de las ecuaciones de Maxwell. Los resultados de la teoría de Hertz se asemejan en muchos aspectos a los de Heaviside, quien además estaba dispuesto a aceptar un término adicional en las ecuaciones de Maxwell que involucrará a *rot*(***V***×***E***), al que ellos llamaron corriente de convección dieléctrica.

Sin embargo, los contemporáneos de Hertz no aceptaron como totalmente acertadas sus tentativas de ampliar la teoría de campo electromagnético al caso en el cual los cuerpos ponderables están en movimiento. Los supuestos de Hertz sobre el éter eran conceptualmente contrarias con la interpretación existente del experimento de Fizeau y parecía discrepar con las fórmulas de Fresnel. Mientras tanto, en la década de 1890, la electrodinámica de medios en movimiento fue tratada sistemáticamente por Lorentz. Las diferencias principales por las cuales la teoría desarrollada por Lorentz se diferenciaba de la de Hertz, consistían en el concepto del "electrón" y en el modelo del éter. Lorentz diseñó sus ecuaciones de acuerdo con la exitosa teoría de Fresnel. Al suponer que un cuerpo ponderable en movimiento no puede comunicar su movimiento al éter que lo rodea, se hacía una distinción entre la materia y el éter. Esta hipótesis correspondía al éter en reposo e implicaba que ninguna parte del éter podría estar en movimiento respecto a cualquier otra parte. La validez de la hipótesis de Lorentz, en comparación con la de Hertz, fue confirmada luego por varios experimentos, uno de cuales fue el descubrimiento del electrón. La confirmación experimental de que las cargas eléctricas residen en los átomos puso la teoría de Lorentz en el centro del interés científico y la convirtió en la base de todas las investigaciones en el área del electromagnetismo. La reconciliación de la teoría electromagnética con la ley de la propagación de la luz en los cuerpos móviles de Fresnel, alcanzado en el marco de la teoría de Lorentz fue considerado como un nuevo logro. Sin embargo, la teoría estaba lejos de resolver completamente algunos problemas planteados. Algunos de ellos eran internos, otros estaban relacionados con la aplicación directa de la teoría.

La consolidación decisiva de la teoría de Lorentz de cuerpos en movimiento fue hecha por A. Einstein quien estableció un principio fundamental de relatividad. Desde entonces se dispone de un esquema más comprensivo y consistente de los fenómenos electromagnéticos. Esto fue un punto decisivo en el desarrollo del electromagnetismo. El concepto mecánico del éter había sido abolido. Las fórmulas de Fresnel se podían entender ahora en base a la ley relativista de la adición de velocidades. Los argumentos principales contra la teoría de Hertz fueron eliminados. Sin embargo, la teoría de Lorentz, conceptualmente renovada, (sin cualquier tentativa de hacer lo mismo con la teoría de Hertz) se mantuvo como la base de la electrodinámica clásica convencional. A pesar de todos los avances, la teoría electromagnética de Lorentz tiene inconsistencias internas tales como la fuerza de auto-reacción (auto-interacción), la contribución infinita de la auto-energía, el concepto de la masa electromagnética, la indeterminación en el flujo de la energía electromagnética, la uni-direccionalidad de los fenómenos de la radiación con respecto a la reversibilidad de movimiento en las ecuaciones básicas de Maxwell, etc.. El advenimiento de la mecánica cuántica a principios de este siglo trajo la esperanza de que todas las dificultades clásicas se podrían rectificar en los marcos de la electrodinámica cuántica. La teoría cuántica de campo alivia algunos problemas pero no puede superarlos sin introducir los injustificados métodos de renormalización. Las dificultades principales en la teoría de Maxwell persisten y no desaparecen a pesar de considerar las modificaciones de la mecánica cuántica. La teoría de Hertz, renovada de acuerdo con el principio de relatividad de Einstein, pudo proporcionar una alternativa más satisfactoria.

Como se demuestra en [13] una modificación de las ecuaciones de Maxwell-Hertz se puede hacer equivalente a una forma del sistema de ecuaciones de Maxwell-Lorentz, y su solución rigurosa demuestra la existencia de una componente longitudinal del campo electromagnético que depende explícitamente del tiempo. Este hallazgo apoya algunas investigaciones recientes en la nueva área de soluciones longitudinales de las ecuaciones fundamentales de campo electromagnético propuestas por M. W. Evans y J.-P. Vigier [10, 11] y por H. A. Múnera y O. Guzmán [12].

La solución general de nuestras ecuaciones modificadas de campo reproduce la teoría de la "separación de potenciales" propuesta recientemente en [13] como método de eliminar las dificultades antedichas de la teoría electromagnética clásica. En [16] se demuestra que el concepto convencional de campo de Faraday-Maxwell no es completamente adecuado para las ecuaciones de Maxwell. Se desarrolló el dualismo electrodinámico, el cual implica la coexistencia de las interacciones instantáneas de largo alcance y las de corto alcance de Faraday-Maxwell. Por primera vez desde el descubrimiento experimental de Hertz, se ha desarrollado un argumento a favor de la teoría alternativa de Helmholtz renovada en el marco de las ecuaciones de Maxwell. El concepto de acción a distancia en esta teoría difiere totalmente de la acción a distancia en las teorías de relatividad post-especial [14, 15], teorías que suponen que solamente la acción retardada a distancia con la velocidad de la luz puede ser consistente con la relatividad. El nuevo enfoque desarrollado en [16] demuestra la compatibilidad de la acción instantánea a distancia con la electrodinámica clásica relativista. El nuevo "concepto del dualismo" puede tender un puente entre la física clásica y la cuántica. Desde este punto de vista, la acción instantánea a distancia puede convertirse en la analogía clásica de la no-localidad de las teorías cuánticas.

## *II. EL GRUPO DE LORENTZ, ESPINES ALTOS Y PARTÍCULAS NEUTRAS.*

Otro tema de interés es la aplicación de la teoría de grupos de Lorentz y de Poincaré en la derivación de las ecuaciones relativistas con sus correspondientes consecuencias fenomenológicas. Es importante mencionar que la situación con la consideración de simetrías discretas (tales como *C*, *P* y *T*) está lejos de la perfección, particularmente para campos con espines altos $s > 1/2$. Hay otros problemas (como la propagación causal) en la descripción de campos con altos espines. Otro problema importante de la física moderna, la explicación del número de generaciones de quarks y leptones y sus espectros de masa respectivos, todavía no ha sido resuelto. Estos son los temas de interés del Dr. V. Dvoeglazov.

Nos permitimos enlistar sus resultados recientes:

- Derivamos la ecuación para las partículas neutras en las representaciones $(1/2,0)\oplus(0,1/2)$ y $(1,0)\oplus(0,1)$, [17]. Por ejemplo, para los espinores auto/contra-auto conjugados tenemos:

$$i\gamma_\mu\partial_\mu\lambda - m\rho = 0.$$

- Ha sido encontrado el mapeo entre el formalismo de Weinberg, Tucker y Hammer (WTH) para espín 1,

$$[\gamma_{\alpha\beta}p_\alpha p_\beta + A\, p_\alpha p_\alpha + Bm^2]\psi_6 = 0$$

y los campos antisimétricos de segundo rango de los cuatro tipos (bajo la condición de que las soluciones de las ecuaciones WTH son eigen-soluciones de paridad). La correspondencia es:

$$\partial_\alpha\partial_\mu F_{\mu\beta} - \partial_\beta\partial_\mu F_{\mu\alpha} + ((A-1)/2)\partial_\mu\partial_\mu F_{\alpha\beta} - (B/2)m^2 F_{\alpha\beta} = 0,$$

$$\partial_\alpha\partial_\mu F_{\mu\beta} - \partial_\beta\partial_\mu F_{\mu\alpha} - ((A+1)/2)\partial_\mu\partial_\mu F_{\alpha\beta} + (B/2)m^2 F_{\alpha\beta} = 0.$$

También, se puede derivar las ecuaciones para el tensor dual de $F_{\alpha\beta}$. En el caso de Tucker y Hammer ($A=1$, $B=2$) la primera ecuación nos proporciona la teoría de Proca $\partial_\alpha\partial_\mu F_{\mu\beta} - \partial_\beta\partial_\mu F_{\mu\alpha} = m^2 F_{\alpha\beta}$.

- Sus límites sin masa contienen soluciones adicionales, en comparación con las ecuaciones de Maxwell. Esto está relacionado con la existencia teórica de los campos de Ogievetskii-Polubarinov-Hayashi-Kalb-Ramond [18].

- En casos particulares (A=0, B=1) las soluciones de diferentes paridades (con masa) se dividen naturalmente en los tipos de soluciones causales y taquiónicas.

- Si queremos tomar en cuenta las soluciones de las ecuaciones WTH con propiedades con respecto de paridad diferentes, esto nos induce a generalizar los formalismos de Bargmann y Wigner (de tal manera que incluyan los operadores de signo en las correspondientes ecuaciones de Dirac [19]), de Proca, y de Duffin, Kemmer y Petiau.

- En las representaciones del tipo $(s,0)\oplus(0,s)$ es posible introducir una estructura con violación de paridad. Las soluciones correspondientes son la mezcla de diferentes estados de polarización.

- La suma de las ecuaciones en las representaciones $(s,0)\oplus(0,s)$ con la ecuación de Klein y Gordon puede cambiar el contenido teórico hasta en el nivel de los campos libres. Por ejemplo, las ecuaciones nuevas para espines altos pueden describir los estados de espín variable y masa variable.

- El mapeo entre las soluciones de WTH de paridad indefinida y los campos antisimétricos existe. En este caso tenemos que introducir el tensor dual. Son ocho.

- En el trabajo anterior [20] Dvoeglazov presentó la teoría de los campos $(1/2,0)\oplus(0,1/2)$ en la base de helicidad. Con respecto de las inversiones de espacio los diferentes estados de helicidad transforman uno a otro, $P\,u_h(-\mathbf{p}) = -i\,u_{-h}(\mathbf{p})$, $P\,v_h(-\mathbf{p}) = +i\,v_{-h}(\mathbf{p})$.

- La representación $(1/2,1/2)$ contiene los estados de espín 1 (bien conocido) y el de espín 0 (compare con el formalismo de Stückelberg). Si no tomamos el último, el conjunto de 4-vectores *no* forman un sistema completo en sentido matemático.

- No podemos eliminar los términos $(\partial_\mu B^*_\mu)(\partial_\nu B_\nu)$ del Lagrangiano y de los invariantes dinámicos si no aplicaríamos el método de Fermi (en otras palabras, "manualmente"). La condición de Lorentz se aplica solamente a los estados de espín 1.

- Tenemos los términos adicionales en las expresiones para el vector de energía y momento, de 4-corriente y del vector de Pauli y Lubanski. Son consecuencia de la imposibilidad de aplicar la condición de Lorentz para los espines $s=0$.

- Si escogemos los 4-potenciales y los campos electromagnéticos [21,22] en base de helicidad los vectores de polarización tienen diferentes propiedades con respecto a la operación de paridad en comparación con la elección de base estándar. Los eigenvectores de helicidad ($s=1$) *no* son los eigenvectores de paridad como en el caso de la representación $(1/2,0)\oplus(0,1/2)$. Sin embargo, la paridad es un numero cuántico "bueno" $[P,H]\_=0$ en el espacio de Fock [22].

- Podemos describir los estados de diferente masa a partir de primeros principios en la representación $(1/2,1/2)$.

- Los operadores de campo pueden ser construidos de manera diferente. Pueden contener los estados *C*, *P* y *CP* conjugados. Aunque, es posible que $b_\lambda^\dagger = a_\lambda^\dagger$, podemos tener los campos complejos de 4-potencial.

- En la teoría generalizada de Stückelberg los propagadores se comportan adecuadamente en el límite sin masa. Esto es opuesto a la teoría de Proca. Lo anterior nos permite explicar las razones físicas para la introducción de la métrica indefinida.

## *III. EL INCREÍBLE NEUTRINO.*

Los neutrinos son unas de las partículas fundamentales las cuales constituyen el universo. Ellos son también los menos conocidos.

Las características poco comunes del neutrón, especialmente su desintegración, llevaron a la introducción y a la posterior detección de una nueva partícula, el neutrino.

Antes de postular la existencia del neutrino, se pensaba que los productos del decaimiento beta solamente consistían en: n→p+e⁻ un protón con la emisión de un electrón, lo que traía como consecuencia la violación de las leyes de conservación de la energía, del momento lineal y del momento angular.

Posteriormente Wolfang Pauli en 1931 propuso la existencia del neutrino con el fin de asegurar la ley de conservación de la energía del momento lineal y del momento angular en los decaimientos beta. Pauli supuso que durante cada desintegración beta, se emite no una, sino dos partículas, es decir, además del electrón se emite una partícula más sin carga de masa muy pequeña y de espín $\hbar/2$. Así Enrico Fermi basándose en esta hipótesis construyó los principios de la moderna teoría de la desintegración beta. Propuso llamar neutrino a la nueva partícula postulada por Pauli, lo que significa pequeña partícula neutra.

La partícula introducida por Pauli tiene una alta propiedad de penetración y es más inatrapable que los rayos gamma. Esta partícula se lleva consigo parte de la energía, del momento lineal y del momento angular.

La confirmación experimental del neutrino se efectuó hasta 1956 con los experimentos de C. L. Cowan y F. Reines, los cuales recibieron el Premio Nóbel de Física 30 años más tarde.

Los neutrinos no tienen carga eléctrica, ellos son eléctricamente neutros, por lo cual no son afectados por la fuerza electromagnética la cual actúa sobre electrones. Los neutrinos son afectados solamente por una fuerza subatómica "débil" de rango mucho más corto que el electromagnetismo, y son por lo tanto capaces de pasar a través de grandes distancias en materia sin ser afectados por ella. Si los neutrinos tienen masa, ellos también interactúan gravitacionalmente con otras partículas masivas, sin embargo la gravedad es por mucho la más débil a nivel microscópico de las cuatro fuerzas fundamentales. Una opinión diferente se encuentra en [43].

Tres tipos de neutrinos son conocidos hasta ahora, y hay fuertes evidencias de que no existen neutrinos adicionales a menos que sus propiedades sean inesperadamente muy diferentes de los tipos conocidos. Cada tipo o "sabor" de neutrino está relacionado con una partícula cargada (la cual da al correspondiente neutrino su nombre). Por lo tanto, el "neutrino del electrón" es asociado con el electrón, y los otros dos neutrinos con el muón y el tau respectivamente. La siguiente tabla enlista los tipos conocidos de neutrinos y sus parejas eléctricamente cargadas:

| Neutrino | $\nu_e$ | $\nu_\mu$ | $\nu_\tau$ |
|---|---|---|---|
| Pareja cargada | Electrón (e) | Muón ($\mu$) | Tau ($\tau$) |

El Modelo Estándar (ME) [23] de la física de las partículas elementales, describe los componentes fundamentales de materia y sus interacciones. Sus bloques básicos son tres generaciones de partículas asociados en 6 quarks y 6 leptones y sus correspondientes portadores de interacción como: gluones, fotones y bosones vectoriales $W^+$, $W^-$ y $Z^0$. Estos bloques son suficientes para formar todo lo existente en el universo, desde un átomo hasta una galaxia, además los datos experimentalmente disponibles hasta ahora para los procesos electrodébiles son bien entendidos en el contexto de este modelo.

Sin embargo, a pesar de su funcionalidad y éxito, durante toda la vida de este modelo se le han hecho múltiples arreglos, sobre todo cuando un fenómeno escapa de su comprensión. Como es el caso del comportamiento del neutrino, cuyos últimos estudios, como los del SUPER-KAMIOKANDE [24] experimento sobre la oscilación de neutrinos, no concuerdan con el ME de un neutrino sin masa, señal inequívoca de que el ME no es completo. Aún más este no es el único experimento en desacuerdo con el modelo, lo mismo es cierto para los experimentos de GALLEX, SAGE, GNO, HOMESTAKE y el Liquid Scintillator Neutrino Detector (LSND) [25].

Por otra parte, en muchas extensiones del ME los neutrinos adquieren una masa diferente de cero, un momento magnético y un momento dipolar eléctrico [26]. De esta manera los neutrinos son excelentes candidatos para poseer características de física más allá del ME [23]. Aparte de sus masas y mezclas, su momento magnético y su momento dipolar eléctrico [27]

| Angulo de mezcla del modelo | Momento magnético | Momento dipolar eléctrico ($10^{-16}$ ecm) |
|---|---|---|
| $\psi$ | $a_\tau$ | $d_\tau$ |
| -0.009 | 0.06 | 3.27 |
| -0.005 | 0.059 | 3.25 |
| 0 | 0.058 | 3.22 |
| 0.004 | 0.057 | 3.21 |

son también señales de nueva física y son de relevancia en experimentos terrestres, como en el problema de los neutrinos solares, en astrofísica y en cosmología [28].

El neutrino, la clave de grandes incógnitas, y quizás la causa de una nueva revolución científica, puede dar luz a múltiples problemas, todos de gran relevancia y posible causa de cambios en la percepción del universo, por ejemplo:

- El problema de los neutrinos solares y su oscilación es uno de los fenómenos para los cuales el modelo estándar no puede dar una explicación.
- El neutrino es un candidato de materia oscura, para explicar la masa faltante del universo.

Hoy en día, en muchos congresos de física de partículas elementales, y en física de astro-partículas a nivel nacional e internacional el tema del neutrino es hoy por hoy el tema más sonado, junto con el problema de la detección del bosón de Higgs del ME. En este momento se llevan a cabo millonarias inversiones con equipos de lo más variado y sofisticado y en lugares y posiciones de lo más peculiares, como en el fondo de una mina de oro ó en la cima de un monte en Italia, con el único fin de saber mucho más acerca de esta partícula, el neutrino.

## *IV. UNA ANALOGÍA ENTRE LA ELECTRODINÁMICA CLÁSICA Y LA TEORÍA GENERAL DE LA RELATIVIDAD.*

El significado físico de las soluciones de las ecuaciones del Einstein está condicionado parcialmente por la determinación de las características dinámicas globales de estos campos, las cuales pueden darnos pistas sobre la naturaleza física de las fuentes que originan los campos gravitacionales descritos por estas soluciones. Sin embargo, la definición de la densidad de estas cantidades, la densidad de la energía-impulso del sistema materia-gravitación, es un problema que no ha sido totalmente entendido debido al enfoque geométrico de la teoría general de la relatividad de Einstein (TGR).

Parece natural que en cualquier análisis concreto en el marco de la TGR siempre sea necesario fijar el sistema de coordenadas. Esto es una consecuencia del hecho de que las ecuaciones de Einstein son ecuaciones para 10 funciones independientes de la métrica, pero debido a las identidades de Bianchi se reducen a sólo seis ecuaciones, lo que significa que hay cuatro grados de libertad. Estas corresponden a la arbitrariedad en la elección del sistema de coordenadas.

Es necesario observar que existe una analogía directa con la definición de los potenciales en la electrodinámica. De hecho, las ecuaciones de Maxwell para los potenciales forman un sistema de cuatro ecuaciones, pero además es necesario considerar la ley de conservación de la carga, lo que da un grado de libertad, que se expresa en la transformación gradiental de los potenciales. Esta indeterminación es eliminada al elegir condiciones de calibración para los potenciales, por ejemplo la calibración de Lorentz.

Si llevamos esta analogía más lejos, uno puede proponer una condición sobre las coordenadas como una expresión análoga a la calibración de Lorentz, tomando como potencial la densidad del tensor métrico contravariante (condición de De Donder-Fock) [35]

$$\left(\sqrt{-g}\, g^{\alpha\beta}\right)_{,\alpha} = 0.$$

Esta condición define el sistema coordenado llamado armónico.

Por lo anterior, si siempre es necesario fijar el sistema coordenado de modo que las ecuaciones de Einstein formen un sistema cerrado, entonces con esto se elimina la restricción, impuesta comúnmente a las magnitudes que definen las características que se conservan para el campo gravitacional, de que deban ser representados mediante cantidades tensoriales. Por esta razón es natural que la densidad de las cantidades que se conservan esté descrita a través de objetos no-tensoriales, que dependen del sistema de coordenadas empleado. La pregunta sobre el significado físico de cada sistema coordenado y cuál de ellos deba considerarse privilegiado, se resuelve parcialmente para los sistemas insulares en el marco del problema de la determinación de las características dinámicas globales, puesto que para estos sistemas es natural elegir coordenadas que se comporten en el infinito como coordenadas cartesianas.

En 1948 Papapetrou [29] propuso para las leyes de conservación una variante simétrica de pseudotensor basada en el estudio del campo gravitacional relacionado con el espacio-tiempo plano, es decir, esencialmente, con el formalismo bimétrico, como fue demostrado claramente después por Burlankov [30].

Siguiendo las ideas de Rosen [31, 32], Papapetrou en su artículo propone interpretar la teoría gravitacional de Einstein, al igual que las otras teorías del campo, como una teoría relacionada al fondo del espacio-tiempo plano, dando a las componentes del tensor métrico el sentido de potencial gravitacional e introduciendo simultánea e independientemente el tensor métrico del espacio-tiempo plano. Se debe observar que Papapetrou en sus trabajos anteriores sobre la ley de la conservación del momento angular en la TGR, llega a la conclusión de que esta ley se debe formular en la teoría de Einstein solamente en el caso de una introducción explícita del tensor métrico del mundo plano. Esta interpretación posee la siguiente ventaja matemática: todas las cantidades que son pseudotensores en la TGR, y todas las relaciones pseudotensoriales, como por ejemplo la ley de la conservación de la energía-momento, adquieren carácter tensorial si la derivada covariante se introduce con respecto a la métrica plana auxiliar.

Como es conocido, en la teoría especial de la relatividad la ley de conservación del momento angular es consecuencia de dos hechos: en primer lugar la invariancia de la integral de acción con respecto a transformaciones de rotación, y en segundo lugar la propiedad de simetría del tensor de energía-impulso de la materia. En el TGR, por un lado, no hay una noción natural de traslación y rotación, y esto no permite deducir la ley de conservación del momento angular directamente del teorema de Noether, y por otro lado la expresión obtenida por este teorema para el complejo energía-impulso no posee la propiedad de simetría.

Sin embargo, la introducción de la métrica del mundo plano, que lleva a un nuevo transporte independiente de la trayectoria, permite aplicar en la TGR conceptos tales como traslación y rotación, y por lo tanto formular la ley de conservación para el momento angular. Del hecho de que la función de Lagrange no-covariante es una densidad escalar respecto a transformaciones lineales (traslación y rotación), se puede obtener la ley de la conservación de la energía-impluso con el pseudotensor canónico para el campo gravitacional y finalmente obtener la ley de conservación del momento angular al introducir la métrica galileana.

Otra expresión para el momento angular se obtiene si requerimos que el complejo energía-impulso sea simétrico. Esta forma de definición fue la que utilizó Papapetrou haciendo uso del método general de simetrización propuesto por Belinfante [33]. Con la expresión simétrica del pseudotensor de Papapetrou se obtiene la correspondiente ley de conservación del momento angular, con la ventaja de que al usar la condición de De Donder-Fock (coordenadas armónicas) esta expresión se simplifica notablemente. La elección del pseudotensor de Papapetrou en coordenadas armónicas para la determinación de las características dinámicas globales (energía, impulso, momento angular y momentos cuadrupolares) es favorable debido al hecho de que todas estas cantidades permiten, en el caso de campos estacionarios, la aplicación del teorema de Gauss-Ostrodgaskii para calcular integrales por superficies bidimensionales, donde los integrandos tienen una forma muy simple [34].

Las coordenadas armónicas fueron introducidos por De Donder y Fock [35]. Para este último el sentido físico de las coordenadas armónicas se reduce a una especie de generalización de las coordenadas cartesianas del mundo plano al caso del espacio-tiempo curvo. Esta generalización contiene dos momentos. Primero, en el mundo plano en un sistema de referencia inercial las coordenadas cartesianas son privilegiadas debido a que las transformaciones de Lorentz, que expresan la homogeneidad y la isotropía del espacio-tiempo, tienen una forma lineal. En el caso del espacio-tiempo homogéneo solamente en el infinito (el caso de sistemas insulares), es posible introducir un sistema de coordenadas privilegiado con una exactitud de hasta las transformaciones de Lorentz. Se entiende que estas coordenadas deben comportarse como cartesianas para distancias infinitamente grandes de donde se ubica la fuente. En segundo lugar, como en el caso de las coordenadas cartesianas, las coordenadas armónicas, de acuerdo con Fock, elimina todos los campos gravitacionales ficticios.

La forma funcional de la métrica, que permite la separación de variables en la ecuación de D'Alambert y por lo tanto también permite encontrar las coordenadas armónicas correspondientes, fue obtenida.en [34]. En calidad de ejemplos de aplicación se realizaron los cálculos para obtener las coordenadas armónicas y las características dinámicas globales para los campos de Schwarzschild y de Kerr. Los resultados obtenidos confirman, en principio, la interpretación física comúnmente asociada a estos sistemas [36-40]. Se puede aplicar este método para la interpretación de otras soluciones estacionarias insulares con simetría axial de las ecuaciones de Einstein tales como las soluciones NUT [41] y el hilo de luz [42].

Finalmente, los autores del presente artículo queremos expresar nuestro deseo de que la lectura de este escrito pueda contribuir a despertar el interés hacia la física entre los jóvenes lectores y de esta manera cumplir con uno de los objetivos planteados por los organizadores de la celebración del Año Internacional de la Física.

## Referencias: